\documentclass{hep99}
\input{epsf}
\begin{document}

\title{Electroweak Precision Measurements\\
 with Leptons}

\author{James E. Brau, for the SLD Collaboration}

\address{Physics Department, University of Oregon, Eugene, OR 97403-1274 USA
\\E-mail: {\tt jimbrau@faraday.uoregon.edu}}

\abstract{The precision measurements of lepton electroweak parameters
at SLD and LEP are reviewed and discussed.  The updated SLD  
weak mixing angle measurement from $A_{LR}$ and the lepton left-right forward-backward asymmetries 
is $\sin^2\theta_W^{eff} = 0.23099 \pm 0.00026$,
and the combined SLD/LEP lepton-based value is $0.23119 \pm 0.00020$.
This value differs by over $3\sigma$ from the comparable quark-based value.
} 

\maketitle

\section{Introduction\label{sec:int}}

The 90's era of precision electroweak measurements at the Z has 
come to an end with the last run of SLD.  SLD has completed the 
analysis of all of its data.  The left-right asymmetry measurement,
the SLD left-right forward-backward asymmetry for leptons, and the
LEP tau polarization measurements have all recently been
updated.  Here we summarize
these recent results.

\section{SLD Left-right Asymmetry\label{sec:alr}}

The SLD left-right asymmetry measurement is unique among all electroweak
precision measurements in that no efficiency or acceptance
corrections are needed.  Furthermore, the final state identification
is relatively unsophisticated.  These features allow a careful measurement
with  a small systematic error (now $\sim 0.65\%$), much smaller than
the statistical error ($\sim 1.3\%$).

The polarized differential cross section at the Z pole is given by :
$$ {d\sigma \over {d cos\theta}} \sim (1 - {\cal P}_e A_e)(1 + \cos^2\theta) 
   + 2A_f(A_e - {\cal P}_e)cos\theta, $$
where the parity violating asymmetries in terms of the
vector and axial vector NC couplings for fermion flavor $f$ 
are $A_f = {2v_f a_f \over {v_f^2 + a_f^2}}$.
The polarized $e^-$ beam at the SLC allows for the isolation of the
initial state ($A_e$) and final state ($A_f$) asymmetries.  The initial state
couplings are determined most precisely via the left-right Z production asymmetry

$$ A_{LR}^0 = {1 \over {\cal P}_e} 
   {\sigma_L - \sigma_R \over {\sigma_L + \sigma_R}} = A_e .$$

The precision of the SLD left-right asymmetry measurement has
improved substantial during the 90's as an increasing luminosity 
has reduced the statistical error, and improved controls and
understanding of the experiment has led to smaller systematic errors.
The unique precision of the left-right asymmetry demands extensive 
cross-checks to confirm the measurement, and three of the
significant recent checks have been the secondary, independent measurements of
the electron polarization, the verification of the center-of-mass collision
energy, and the measurement of the positron polarization (confirming its
non-existence).

Table 1 presents the history of SLD luminosity and polarization.
The experiment was capped with its most productive run in 1997-98, when 
approximately 350 thousand Z bosons were detected.  The peak
luminosities achieved were about $3 \times 10^{30} {\mathrm cm^{-2}s^{-1}}$.

\begin{table}
\begin{center}
\caption{Summary of SLD Runs}
\begin{tabular}{llllll} 
\br
&'92&'93&'94-5&'96&'97-8 \\ 
\mr
Total $Z^0$'s& 10k & 50k & 100k & 50k & 350k \\
$<$Pol$>$ & 22\% &63\% & 78\% & 78\% & 73\% \\ 
\br
\end{tabular}
\end{center}
\label{tab:runs}
\end{table}

\subsection{Evolution of SLD systematic errors}

Table 2 presents the evolution of the SLD systematic errors.
The final achieved systematic error of $0.65\%$  represents several years of instrumental
work and cross-checks, supplemented by extensive accelerator 
based tests.  This effort has established  a high
degree of confidence in the measurement of the left-right
asymmetry.
Three of the recent checks are described below.

\begin{table*}
\begin{center}
\caption{Evolution of the SLD systematic errors}
\begin{tabular}{lllll} 
\br
&1992&1993&1995&Now \\ 
\mr
Polarimetry & 2.7\% &1.7\% & 0.67\% & 0.5\% \\ 
$E_{cm}$& $ \stackrel{<}{\sim} $0.3\% & $ \stackrel{<}{\sim} $0.3\%  & 0.3\% & 0.4\% \\ 
Background (frac.)& 1.4\% & 0.25\% & 0.11\% & 0.044\% \\
SLC asymmetry ($10^{-4}$) & $1.8 \pm 4.2$ & $0.4 \pm 0.5$ &
$-1.9 \pm 0.3$ & $-1.3 \pm 0.7$\\
\mr
Total fractional error & 3.6\% & 1.7\% & 0.75\% & 0.65\% \\
\br
\end{tabular}
\end{center}
\label{tab:syst}
\end{table*}

\subsection{Recent SLD checks}

{\bf Electron polarization.}
Two additional independent measurements have been made to confirm the 
electron polarization measurement.  The primary polarimeter for SLD
is the Cerenkov detector which detects the Compton scattered electrons in a Compton scattering polarimeter
system just downstream of the $e^+e^-$ interaction point.  
It measures the asymmetry in the Compton cross section at the kinematic
edge, with a 70\% analyzing power.

Now, two additional detectors measure the Compton scattered gammas.  
The Polarized Gamma Counter (PGC) consists of a Cerenkov detector behind
a variable thickness of lead radiator.  It has an analyzing power
of 16 to 22 \%, depending on the amount of lead.  The Quartz Fiber
Calorimeter (QFC) absorbs the Compton gammas, with an analyzing
power of 18\%.  These polarimeters have confirmed the Cerenkov measurements.


{\bf Center-of-mass collision energy.}
A scan of the $Z^0$ resonance confirms the SLC beam energy measurements.
Two off energy points were taken and the center of mass energy
was found to be off the true Z peak by $-46 \pm 25 MeV$.

{\bf Positron Polarization.}
A measurement of the positron polarization was made
with the End Station A fixed target polarimeter.
The result was $P_{e^+} = -0.02\% \pm 0.07\%$, consistent with zero.

\subsection{The final SLD result on $A_{LR}$}
Table 3 shows the time history of the SLD results for $A_{LR}^0$.
The measurement has a consistent history, with improving precision,
to the current value of

$$A_{LR}^0 = 0.15108 \pm 0.00218$$

\noindent which translates into a value for $\sin^2\theta_W^{eff}$ of

$$\sin^2\theta_W^{eff} = 0.23101 \pm 0.00028$$

\begin{table}
\begin{center}
\caption{History of the SLD measurement of $A_{LR}$}
\begin{tabular}{ll} 
\br
Year&$A_{LR}^0$ \\ 
\mr
1992 & $0.100 \pm 0.044 \pm 0.004$    \\ 
1993  &   $0.1656 \pm 0.0071 \pm 0.0028$  \\ 
1995  &  $0.1512 \pm 0.0042 \pm 0.0011$ \\
1996    &   $0.1570 \pm 0.0057 \pm 0.0011$ \\
1997-98 &   $0.1490 \pm 0.0024 \pm 0.0010$    \\
\br
\end{tabular}
\end{center}
\label{tab:history}
\end{table}

\section{SLD Left-right Forward-backward Asymmetry for electron, muons, and taus\label{sec:lrfb}}

SLD has updated its left-right forward-backward asymmetry measurement
for the leptons.  This measures the final state coupling
of the Z to leptons.  The asymmetry is defined as

$$\tilde A_{FB}^f = {(N_{LF}-N_{LB})-(N_{RF}-N_{RB}) \over 
(N_{LF}+N_{LB})+(N_{RF}+N_{RB})} = {3 \over{4}} {\cal P}_e A_f,$$
for final state lepton flavor $f = e,\mu ,\tau$  

Each lepton asymmetry, $A_f$ is determined by fitting the angular distribution:

$${{\rm d} \sigma \over {\rm d} cos\theta} \sim
(1 - P_e A_e)(1 + cos^2\theta) + 2 A_{\mu , \tau}(A_e - P_e) cos \theta,$$
as shown in Figure 1.

The combined preliminary results for the 1993-98 data are consistent with 
lepton universality:

$$\matrix{ A_e = 0.1558 \pm 0.0064 \cr
A_{\mu} = 0.137 \pm 0.016\cr
A_{\tau} = 0.142 \pm 0.016} \Biggr\}
A_{e\mu\tau} = 0.1523 \pm 0.0057$$
$$\sin^2\theta_W^{eff} = 0.23085 \pm 0.00073$$

There is analysis remaining to be done on this data.  Presently the
angular coverage is $|\cos \theta| < 0.8$, but will 
be extended to $|\cos \theta| < 0.9$.  With this coverage, 
and with improved efficiencies, the
error on $\sin^2\theta_W^{eff}$ is expected to reach $\pm 0.0006$.

When SLD combines this left-right forward-backward asymmetry
result with the $A_{LR}$ result above,
the SLD measurement for the 
electroweak mixing angle for leptons is:
$$\sin^2\theta_W^{eff} = 0.23099 \pm 0.00026$$

\begin{figure}[b]
\vspace{0cm}
\hspace{0cm}
\centerline{%
\epsfverbosetrue\epsfysize 3.1in 
\epsffile{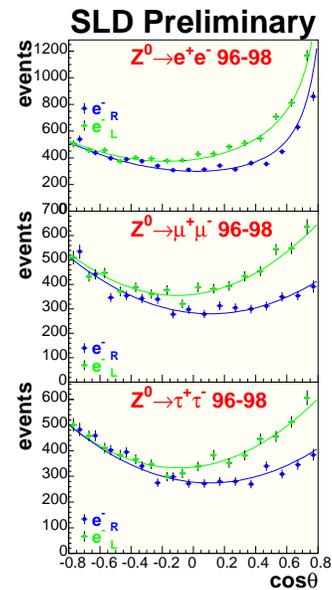}
}
\caption{
SLD di-lepton angular distributions.}
\label{fig:lrfba}
\end{figure}


\section{$\tau$ Polarization from LEP\label{sec:taup}}

There is a new measurement of $\tau$ polarization presented by
DELPHI at this meeting.  The $\tau$ polarization measurements
from LEP now are as presented in Table 4.

\begin{table}
\begin{center}
\caption{$\tau$ polarization results}
\begin{tabular}{lll} 
\br
&$A_\tau$&$A_e$ \\ 
\mr
ALEPH & $.1452 \pm .0061$ & $.1505 \pm .0069$  \\ 
DELPHI& $.1359 \pm .0096$ & $.1382 \pm .0116$ \\ 
L3& $.1476 \pm .0108$ & $.1678 \pm .0130$  \\
OPAL & $.1340 \pm .0134$ & $.1290 \pm .0149$ \\
\br
\end{tabular}
\end{center}
\label{tab:taupol}
\end{table}

\section{Consistency of Electroweak Lepton Measurements}

The combined SLD measurement of $\sin^2\theta_W^{eff}$ from leptons is
$0.23099 \pm 0.00026$.  The LEP measurement of lepton forward-backward
asymmetries, and tau polarization have been combined into
a LEP lepton-based $\sin^2\theta_W^{eff}$ measurement of
$0.23151 \pm 0.00033$.  Combining these two purely lepton
measurements yields:

$$\sin^2\theta_W(world-leptons) = 0.23119 \pm 0.00020.$$
\noindent These results are fully consistent.

When the quark-based measurements are compared to these lepton-based
results (see Figure 2), some indication of inconsistency is suggested.
The hadrons only measurement of $\sin^2\theta_W^{eff}$ is
$0.23236 \pm 0.00031$, more than $3\sigma$ from the lepton-based
measurement

\begin{figure}[b]
\begin{center}
\epsfverbosetrue\epsfysize 2.1in 
\epsffile{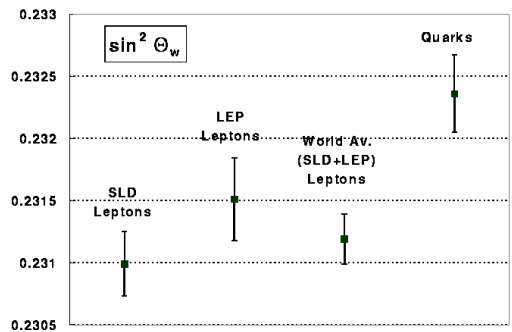}
\end{center}
\caption{Comparison of the lepton- and quark-based measurements of the 
electroweak mixing angle}
\label{fig:pol-comp}
\end{figure}

\section{Conclusion}
SLD has completed the analysis of the data from its final run.
The SLD and LEP measurements of lepton couplings to the Z are fully
consistent, while a significant difference from the quark
sector has been observed.


\end{document}